# A Simple Recommender Engine for
# Matching Final-Year Project Student with Supervisor

Mohammad Hafiz Ismail[1,a], Tajul Rosli Razak[2,], Muhamad Arif Hashim[3],
Alif Faisal Ibrahim[4]

[1,2,3,4]Faculty of Computer and Mathematical Sciences

Universiti Teknologi MARA Perlis, Malaysia

[a]MohammadHafiz@perlis.uitm.edu.my (corresponding author), [a]mypapit@gmail.com

**Abstract.** This paper discusses a simple recommender engine, which can match final year project student based on their interests with potential supervisors. The recommender engine is constructed based on Euclidean distance algorithm. The initial input data for the recommender system is obtained by distributing questionnaire to final year students and recording their response in CSV format. The recommender engine is implemented using Java class and application, and result of the initial tests has shown promises that the project is feasible to be pursued as it has the potential of solving the problem of final year students in finding their potential supervisors.

## Introduction

Final year project (FYP) is one of the requirements to be fulfilled by Bachelor degree students in FSKM, UiTM Perlis in order to graduate. The final year project is divided into two parts: proposal and project construction[1]. The proposal part is crucial as students must determine the project problem area, its significance, scope and objectives. The student may choose his or her own problem area based on their interests or may consult with a potential supervisor from a pool of faculty lecturers to determine a suitable problem area. Regardless of the choice, the student may have to select a supervisor that best match his or her area of interests.

However, as in the current situation, almost all final year students are not familiar with faculty lecturers and their area of expertise as they are only familiar with lecturers who have taught them in previous semester. This will largely affect student decision in selecting project title as they may perceive to have only a limited choice of potential supervisors to pick from and may end up with project titles that are not aligned with their own interests. Furthermore, the students themselves are still new with the concept of research may benefit by having discussion with several lecturers to determine the project area that best suit them.

The situation can be remedied if the students have sufficient information on potential supervisors and their area of expertise. Additionally, students also can benefit in knowing which potential supervisors that have similar interests with each other so that the student can focus in selecting supervisors from a pool of lecturers with similar research interests.

Thus, we proposed a recommender engine for matching FYP students with potential supervisors according to their area of interests. The system would be able to accept the input from the student, which gauge their area of interests and return a list of lecturers that best match the student input.

## Background

### *Recommender System*

Recommender systems are software tools and techniques that provide suggestions for users [2]. Recommender System is typically used in decision-making process and provides users with a list of items that are similar or have relationship with each other. For the purpose of this study, we have implemented a simple recommender engine to suggest a list of lecturers based on students' interest. The recommender engine is implemented using Java class and is designed to be easily integrated into any types of Java application that uses Java 1.6 specification and later (which includes Android mobile application). There are several algorithms used in recommender system, but for this particular study, we chose to use Euclidean distance score to demonstrate our simplified recommender engine.

### *Euclidean Distance Score*

Euclidean distance is described as a distance between two points which represents variable values (x1, y1) and (x2, y2) [3]. Euclidean distance is an algorithm that is useful because it is equivalent to the distance of objects measured in real world [4] . The formula for Euclidean distance is given in Figure 1, where $p$ is data points, $d$ is the distance function, $S_p$ is the component value of the data points.

$$d(p_1, p_2) = \sqrt{\sum_{i \in \text{item}} (s_{p_1} - s_{p_2})^2}$$

*Figure 1: Euclidian distance formula*

While frequently represented in two-dimensional or three-dimensional space, Euclidean distance is also capable of calculating distance within multi-dimensional variables. The Euclidean distance between points can also be transformed to measure similarity with a simple formula (Figure 2)

$$\frac{1}{1 + d(p_1, p_2)}$$

*Figure 2: Euclidean distance to similarity score conversion formula*

Although not required, the conversion to similarity score will give result between 0.0 and 1.0, where the value near 1.0 represents complete similarity and value near 0.0 represent completely dissimilar item. This is in contrast with measure distance of distance where the 0.0 represents similar item and a higher number (unbounded) would represents item that are not similar [5].

## Methodology

The methodology for this study is divided into four phases as outlined in Figure 3.

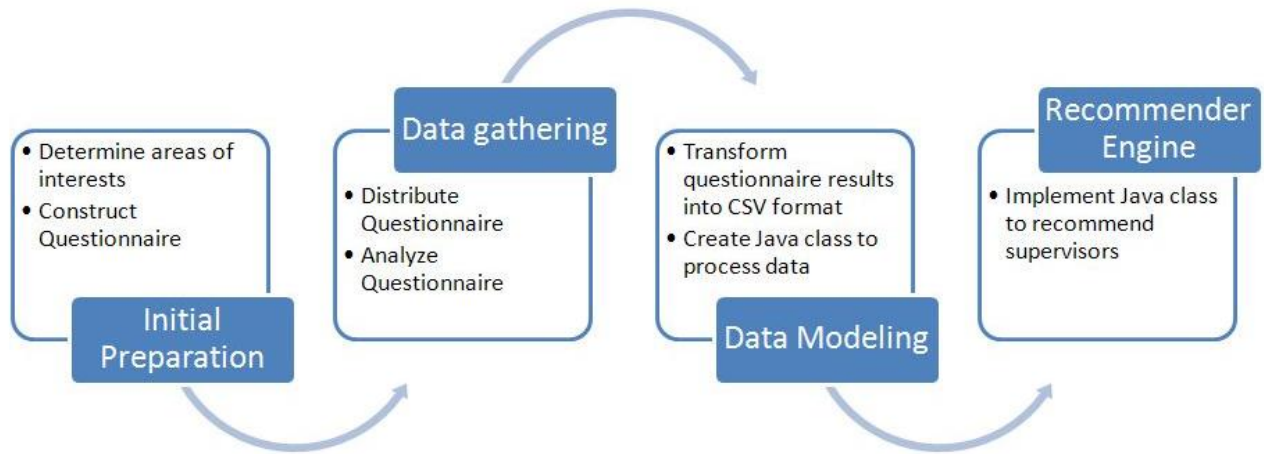

*Figure 3: Recommender engine for matching supervisor methodology*

### Initial Preparation

In this phase, we have studied gathered the previous and current project titles associated with each supervisor and established that the titles falls within four general areas: Multimedia, Web Application, Network, Artificial Intelligence and Mobile Application.

Based on these findings, we have constructed a questionnaire that asks the respondents to rate their interest in a scale of 1 (least interested) to 5 (most interested) in each area. The respondents are also required to fill other particulars such as their chosen project titles, supervisors and co-supervisors.

### Data Gathering

The questionnaires are distributed among Bachelor in IT and Bachelor in Computer Science (Netcentric Computing) final year students as the respondents where 51 out of 53 of the questionnaires are returned. The questionnaires responses are then transferred into Microsoft Excel sheet, where the data is arranged in rows based on student area of interests and supervisors that they have chosen.

### Data Modeling

A Java class named Lecturer is created to model the data based on the area of interests. The Lecturer object represents a single supervisor. The data gathered from the previous phase is transformed into comma separated value format (CSV) and is recorded into a vector of Lecturer objects. Sample format of the CSV file is shown in Figure 4.

Function to calculate method and similarity between the Lecturer objects is implemented using Euclidean distance as *calculateDistance()* and *getSimilarity()* Java method.

```
Abdul Hapes bin Mohamed,5,3,3,4,5
Ahmad Yusri Dak,4.5,1.5,1.5,1.5,3.5
Alif Faisal,5,5,3,3,5
Arifah Fasha bt Rosmani,4.5,4.5,2.5,3,3
Arzami bin Othman,4,4,1,2,3
Dr Ahmad Hanif Baharin,4.5,4.5,3.5,4,4.5
Dr Shukor Sanim bin Mohd Fauzi,4,4,4.5,4,4
Faris,4,3,4,3.5,4.5
Hanisah Ahmad,4,3,1,2,2
Hawa bt Mohd Ekhsan,3.5,4,2,3.5,4.5
IMAN HAZWAN,3.5,4.5,3,4,4.5
Jiwa Noris Hamid,2.5,2,0,0,5
Mahfudzah Othman,4,3.5,3.5,3.5,3
```

*Figure 4: Sample CSV data*

**Recommender Engine**

The Java class is then implemented as a Java application to create a rudimentary recommender engine, which accepts rating of each interest area (Multimedia, Web application, Network, Artificial Intelligence and Mobile Application) and returns a list of recommended lecturers based on interest of interest.

For example:

Student X, who are strongly interested in Multimedia and Web Application but less inclined to do a project in Network and Artificial Intelligence wants to find suitable supervisors would enter the following rating to the recommender engine:

Multimedia – **5.0**, Web Application – **4.5**, Network – **1.0**, AI – **2.5**, Mobile Application – **3.0**

Based on the student input, the recommender system will return a list of lecturer based on Table 1, the sample raw data from Java application is shown in Figure 5.

*Table 1: Sample Recommender Engine Results*

| Potential Supervisors | Similarity |
|---|---|
| Arzami | 44.95 |
| Arifah Fasha | 37.62 |
| Nora Yanti | 34.83 |
| Hanisah Ahmad | 32.04 |
| Mohd Nizam Osman | 29.21 |

```
Prob...  @ Jav...  Decl...  Search  Con...  Log...  Log...

Nama:Arzami bin Othman  Distance: 44.94897427831781
Nama:Arifah Fasha bt Rosmani   Distance: 37.61785115301142
Nama:Nora Yanti bt Che Jan     Distance: 34.83314773547883
Nama:Hanisah Ahmad      Distance: 32.03772410170407
Nama:Mohd Nizam Osman   Distance: 29.20697289657773
```

*Figure 5: Sample Raw output from Java application*

**Conclusion**

In this paper, we have demonstrated a simple recommender engine which can be used for recommending supervisors to final year project students using Euclidean distance algorithm. Although it is shown that Euclidean distance seems to achieve the goal for recommending lecturers, the engine is still far from complete as is only designed as a proof-of-concept. The Euclidean distance algorithm used in this recommender engine can only determine the differences between values but could not be used to reliably determine whether a group of Lecturers has consistently similar interests.

Planned future works for this project includes adding more area-of-interests to cover network sub area and integrating a better algorithm to handle bias within data. One of the algorithm going to be implemented in the near future is Pearson correlation algorithm, because it does not rely solely on differences on value, but rather whether the value in variables is consistent with each other.